\documentclass[floats, prd, eqnum, showpacs, nofootinbib, twocolumn, eqsecnum]{revtex4-1}

\usepackage{color,graphicx}
\usepackage{amsfonts}
\usepackage{amssymb}
\usepackage{comment}
\begin{document}

\title{Affine perturbation series of the deflection angle of a ray near the photon sphere of a Reissner-Nordstr\"{o}m black hole} 
\author{Naoki Tsukamoto${}^{1}$}\email{tsukamoto@rikkyo.ac.jp}

\affiliation{
${}^{1}$Department of General Science and Education, National Institute of Technology, Hachinohe College, Aomori 039-1192, Japan \\
}

\begin{abstract}
We investigate the affine perturbation series of the deflection angle of a ray near the photon sphere of an extreme Reissner-Nordstr\"{o}m black hole.
We compare the $0$th and $1$st orders of the affine perturbation series 
with the deflection angle in a strong deflection limit.
We conclude that the $0$th order of the affine perturbation series
is more accurate than the deflection angle in the strong deflection limit
and that we should improve a strong-deflection-limit analysis by using the $0$th order of the affine perturbation series.
\end{abstract}
\maketitle

\section{Introduction}
Recently, LIGO Scientific and Virgo Collaborations have reported the detection of gravitational waves from a black hole binary~\cite{Abbott:2016blz};
Event Horizon Telescope Collaboration~\cite{Akiyama:2019cqa,EventHorizonTelescope:2022xnr} have reported the detection of the shadows of black hole candidates 
in the center of galaxies; and  
Wilkins \textit{et al.}~\cite{Wilkins:2021yst} have reported the detection of x-ray echoes caused by a strong gravity near a black hole candidate.
Theoretical investigations on the strong gravity due to black holes will be important as the details of the black hole systems are observed.

Compact objects such as black holes and wormholes have a photon sphere~\cite{Perlick_2004_Living_Rev,Claudel:2000yi,Perlick:2021aok,Hod:2017xkz,
Sanchez:1977si,Decanini:2010fz,Wei:2011zw,Hasse_Perlick_2002,Koga:2018ybs,Koga:2020gqd,Ames_1968,Synge:1966okc,Yoshino:2019qsh} 
formed by unstable circular photon orbits due to the strong gravity.
The image of rays reflected near the photon sphere in a Schwarzschild spacetime was studied by Hagiwara in 1931~\cite{Hagihara_1931} 
and by Darwin in 1959~\cite{Darwin_1959},
and the image by the photon sphere has been revived several times~\cite{Atkinson_1965,Luminet_1979,Ohanian_1987,Nemiroff_1993,Virbhadra:1998dy,
Frittelli_Kling_Newman_2000,Virbhadra_Ellis_2000,Bozza_Capozziello_Iovane_Scarpetta_2001,Bozza:2002zj,Virbhadra:2002ju,Holz:2002uf,
Perlick:2003vg,Virbhadra:2008ws,Bozza_2010,Tsukamoto:2016zdu,Gralla:2019xty,Shaikh:2019jfr,Shaikh:2019itn,Tsukamoto:2020uay,Tsukamoto:2020iez,Paul:2020ufc,Guerrero:2022qkh,Virbhadra:2022ybp}.
Bozza expressed the deflection angle $\alpha_\mathrm{SDL}$ of a light ray reflected by the photon sphere in a strong deflection limit $b\rightarrow b_\mathrm{m}+0$, 
where $b$ and $b_\mathrm{m}$ are the impact parameter and the critical impact parameter of the ray, respectively, in the following form:    
\begin{eqnarray}\label{eq:sl}
\alpha_\mathrm{SDL}=
-\bar{a} \log \left( \frac{b}{b_\mathrm{m}}-1 \right) + \bar{b} +O(b-b_\mathrm{m}),
\end{eqnarray}
where $\bar{a}$ and $\bar{b}$ were obtained as $\bar{a}=1$ and $\bar{b}=\log \left[ 216(7-4\sqrt{3}) \right]-\pi$ in the Schwarzschild spacetime~\cite{Bozza:2002zj}.
Note that the same deflection angle was obtained by using a radial coordinate by Darwin~\cite{Darwin_1959} and by Bozza~\textit{et al.}~\cite{Bozza_Capozziello_Iovane_Scarpetta_2001}.

Eiroa~\textit{et al.} investigated a numerical method to obtain $\alpha_\mathrm{SDL}$ in an asymptotically-flat, static, and spherically symmetric spacetime with a photon sphere~\cite{Eiroa:2002mk}.
In Ref.~\cite{Bozza:2002zj}, Bozza investigated the deflection angle $\alpha_\mathrm{SDL}$ in the strong deflection limit $b\rightarrow b_\mathrm{m}+0$, 
in a general asymptotically-flat, static, and spherically symmetric spacetime with the photon sphere.
In Bozza's method,  
$\bar{a}$ is obtained in an analytical expression, 
but $\bar{b}$ is often not obtained in an analytical form because a term cannot be integrated analytically.
The analytical form of $\bar{b}$ has been obtained only in a few spacetimes:
The Schwarzschild spacetime~\cite{Bozza_Capozziello_Iovane_Scarpetta_2001,Bozza:2002zj},
a braneworld black hole spacetime~\cite{Eiroa:2004gh}, and five-dimensional and seven-dimensional Tangherlini spacetimes~\cite{Tsukamoto:2014dta}
were obtained by using Bozza's method.

The strong-deflection-limit analysis was extended by several authors and was applied to many spacetimes and lens configurations~\cite{Tsukamoto:2016zdu,Shaikh:2019jfr,Shaikh:2019itn,Tsukamoto:2020uay,Tsukamoto:2020iez,Paul:2020ufc,Bozza:2002af,Eiroa:2002mk,Petters:2002fa,Eiroa:2003jf,Bozza:2004kq,Bozza:2005tg,Bozza:2006sn,Bozza:2006nm,Iyer:2006cn,Bozza:2007gt,Tsukamoto:2016qro,Ishihara:2016sfv,Tsukamoto:2016oca,Tsukamoto:2016jzh,Tsukamoto:2017edq,Hsieh:2021scb,Aldi:2016ntn,Tsukamoto:2020bjm,Takizawa:2021gdp,Tsukamoto:2021caq,Aratore:2021usi,Bisnovatyi-Kogan:2022ujt,Tsupko:2022kwi}.
Tsukamoto extended Bozza's method~\cite{Bozza:2002zj} to apply directly to ultrastatic spacetimes 
with the constant norm of a time-translational Killing vector~\cite{Tsukamoto:2016qro,Tsukamoto:2016zdu}~\footnote{We can apply Bozza's method 
to the ultrastatic spacetime indirectly~\cite{Bhattacharya:2019kkb}.} 
and claimed that the error term $O(b-b_\mathrm{m})$ in Eq.~(\ref{eq:sl}) should be read as $O\left( ( b/b_\mathrm{m}-1) \log ( b/b_\mathrm{m}-1 ) \right)$.
By using the alternative method, the analytic expressions of $\bar{a}$ and $\bar{b}$ for 
an Ellis-Bronnikov wormhole~\cite{Tsukamoto:2016qro,Tsukamoto:2016zdu} and a spatial Schwarzschild wormhole~\cite{Tsukamoto:2016zdu} were obtained. 
Tsukamoto generalized the alternative method to a general asymptotically-flat, static, and spherically symmetric spacetime with the photon sphere~\cite{Tsukamoto:2016jzh}. 
The analytic expressions of $\bar{a}$ and $\bar{b}$ in a Reissner-Nordstr\"{o}m spacetime~\cite{Tsukamoto:2016oca,Tsukamoto:2016jzh},
in a charged black hole spacetime~\cite{Badia:2017art},
and in Kerr and Kerr-Newman spacetimes~\cite{Hsieh:2021scb}, were calculated by the method.

Iyer and Petters~\cite{Iyer:2006cn} considered 
the affine perturbation series of the deflection angle of the ray near the photon sphere of the Schwarzschild black hole in the following form:
\begin{eqnarray}\label{eq:al}
\alpha&=&
\left( \sigma_0+\sigma_1 b_\mathrm{p}+\sigma_2 b_\mathrm{p}^2 +\sigma_3 b_\mathrm{p}^3 +\cdots  \right) \log \left( \frac{\lambda_0}{b_\mathrm{p}} \right) \nonumber\\
&&+\rho_0+\rho_1 b_\mathrm{p}+\rho_2 b_\mathrm{p}^2 +\rho_3 b_\mathrm{p}^3 +\cdots, 
\end{eqnarray}
where $b_\mathrm{p}$ is defined as
\begin{equation}
b_\mathrm{p} \equiv 1-\frac{b_\mathrm{m}}{b},
\end{equation}
where $\sigma_0$, $\sigma_1$, $\sigma_2$, $\sigma_3$, $\lambda_0$, $\rho_0$, $\rho_1$, $\rho_2$, and $\rho_3$ are coefficients.
The affine perturbation series of the deflection angle 
close the gap that developed between the exact deflection angle and the one in the strong deflection limit.
They showed that the $0$th order of the affine perturbation series $\alpha_\mathrm{0th}$ defined by
\begin{eqnarray}\label{eq:0th}
\alpha_\mathrm{0th}&\equiv&
\sigma_0 \log \left( \frac{\lambda_0}{b_\mathrm{p}} \right) +\rho_0
\end{eqnarray}
does not correspond with the deflection angle $\alpha_\mathrm{SDL}$ in the strong deflection limit by Darwin~\cite{Darwin_1959} and by Bozza~\cite{Bozza:2002zj} 
and that the $0$th order of the affine perturbation series $\alpha_\mathrm{0th}$ is more precise than $\alpha_\mathrm{SDL}$.

The Reissner-Nordstr\"{o}m spacetime, which is obtained as an electrovacuum solution of Einstein equations, 
is often used as the toy model of compact objects with a richer structure than the Schwarzschild spacetime
since it can be tractable in analytical calculations.
The shadow image~\cite{deVries:2000,Takahashi:2005hy,Zakharov:2014lqa,Akiyama:2019cqa,Akiyama:2019eap,Kocherlakota:2021dcv},
gravitational lensing~\cite{Eiroa:2002mk,Bozza:2002zj,Eiroa:2003jf,Bin-Nun:2010exl,Bin-Nun:2010lws,Tsukamoto:2016oca,Tsukamoto:2016jzh}, 
and time delay~\cite{Sereno:2003nd} in the Reissner-Nordstr\"{o}m spacetime were investigated. 

In 2002, Eiroa \textit{et al.} applied their numerical method to a Reissner-Nordstr\"{o}m black hole
with a mass $M$ and an electrical charge $Q$ and obtained numerical values corresponding to $\bar{a}$ and $\bar{b}$~\cite{Eiroa:2002mk}.
In the same year, Bozza obtained the analytical expression of $\bar{a}$ 
and the numerical expression of $\bar{b}$~\cite{Bozza:2002zj}.
The analytical form of $\bar{b}$, by using a series on a small charge $Q/M<1$, was also obtained in Ref.~\cite{Bozza:2002zj}.
In 2016, by using the alternative method~\cite{Tsukamoto:2016oca,Tsukamoto:2016jzh}, 
the analytical expression of $\bar{b}$ was obtained  without the series on the small charge.
The deflection angles of the rays in the Reissner-Nordstr\"{o}m naked singularity spacetime 
with the photon sphere in strong deflection limits $b\rightarrow b_\mathrm{m}\pm0$~\cite{Shaikh:2019itn,Tsukamoto:2021fsz,Tsukamoto:2021lpm},
and with a marginally unstable photon sphere in the strong deflection limit $b\rightarrow b_\mathrm{m}+0$~\cite{Tsukamoto:2020iez},
and the deflection angle in the Reissner-Nordstr\"{o}m naked singularity spacetime without the photon sphere~\cite{Chiba:2017nml}
were also investigated.

In this paper, 
we investigate the affine perturbation expansion of the deflection angle of the ray in analytical calculations near the photon sphere of an extreme charged Reissner-Nordstr\"{o}m black hole.
We discuss the error of the deflection angles 
and we conclude that a strong-deflection-limit analysis for a gravitational lensing can be improved by using $\alpha_\mathrm{0th}$.
We use the units in which the light speed and Newton's constant are unity.

\section{Deflection angle in a Reissner-Nordstr\"{o}m spacetime}
A line element in a Reissner-Nordstr\"{o}m spacetime is given by
\begin{equation}
\mathrm{d}s^2=-A(r)\mathrm{d}t^2+\frac{\mathrm{d}r^2}{A(r)}+r^2 \left(\mathrm{d}\theta^2+\sin^2 \theta \mathrm{d}\phi^2\right),
\end{equation}
where $A(r)$ is given by
\begin{equation}
A(r)\equiv 1-\frac{2M}{r} +\frac{Q^2}{r^2}.
\end{equation}
It has an event horizon at $r=M+\sqrt{M^2-Q^2}$ for $\left|Q\right| \leq M$.
There are time translational and axial Killing vectors~$t^\mu \partial_\mu=\partial_t$ and $\phi^\mu \partial_\mu=\partial_\phi$
because of the stationarity and axisymmetry of the spacetime, respectively.
We can assume $\theta=\pi/2$ and $Q \geq 0$ without loss of generality.

From $k^\mu k_\mu=0$, where $k^\mu\equiv \dot{x}^\mu$ is the wave-number vector of a ray and 
the dot denotes a differentiation with respect to an affine parameter on the orbit of the ray,
the orbit is expressed by 
\begin{equation}\label{eq:motion}
-A(r)\dot{t}^2+\frac{\dot{r}^2}{A(r)}+r^2\dot{\phi}^2=0.
\end{equation}
At the closest distance $r=r_0$, it gives 
\begin{equation}\label{eq:r0}
A_0\dot{t}^2_0=r_0^2\dot{\phi}_0^2.
\end{equation}
Here and hereinafter, quantities with the subscript $0$ denote the quantities at the closest distance.
From the conserved energy of the ray defined by
\begin{equation}
E\equiv -g_{\mu \nu}t^\mu k^\nu=A(r)\dot{t},
\end{equation}
the conserved angular momentum of the ray defined by
\begin{equation}\label{eq:L}
L\equiv g_{\mu \nu}\phi^\mu k^\nu=r^2 \dot{\phi}, 
\end{equation}
and Eq.~(\ref{eq:r0}), 
the impact parameter of the ray is obtained as 
\begin{equation}\label{eq:b}
b(r_0) \equiv \frac{L}{E} = \frac{r_0^2 \dot{\phi}_0}{A_0\dot{t}_0}= \frac{r_0^2}{\sqrt{r_0^2-2Mr_0+Q^2}}.
\end{equation}
Here, we have assumed $L\geq 0$ without loss of generality. 
The equation of the motion of the ray~(\ref{eq:motion}) can rewritten as 
\begin{equation}\label{eq:dotr}
\dot{r}^2+V(r)=0,
\end{equation}
where $V(r)$ is the effective potential of the ray defined by
\begin{equation}
V(r) \equiv E^2 \left( -1+\frac{A(r)}{r^2} b^2 \right).
\end{equation}
The motion of the ray is permitted in a region satisfying $V(r)\leq 0$. 
Let $r_\mathrm{m}$ be the largest positive solution of $V^{\prime}(r)=0$ for $Q< 3M/(2\sqrt{2})$.
We obtain $r_\mathrm{m}$ as
\begin{equation}
r_\mathrm{m} \equiv \frac{3M+\sqrt{9M^2-8Q^2}}{2}.
\end{equation}
When the ray has a critical impact parameter $b=b_\mathrm{m}$ (or $r_0=r_\mathrm{m}$), where 
\begin{equation}
b_\mathrm{m}\equiv b(r_\mathrm{m}) = \frac{r_\mathrm{m}^2}{\sqrt{r_\mathrm{m}^2-2Mr_\mathrm{m}+Q^2}},
\end{equation}
the ray forms an unstable circular light sphere, called a photon sphere, at $r=r_\mathrm{m}$ and 
its effective potential holds $V_\mathrm{m}=V_\mathrm{m}^{\prime}=0$ and $V_\mathrm{m}^{\prime \prime}<0$ for $Q< 3M/(2\sqrt{2})$.
Here and hereinafter, quantities with the subscript m denote the quantities at $r=r_\mathrm{m}$ or $r_0=r_\mathrm{m}$. 
The ray with $b<b_\mathrm{m}$ (or $r_0<r_\mathrm{m}$) goes  inside of the photon sphere while 
the ray with $b>b_\mathrm{m}$ (or $r_0>r_\mathrm{m}$) is scattered on the outside of the photon sphere. 
We assume $b>b_\mathrm{m}$ (or $r_0>r_\mathrm{m}$) and $Q< 3M/(2\sqrt{2})$.

From Eq.~(\ref{eq:motion}), 
the deflection angle of the ray is obtained as
\begin{eqnarray}\label{eq:alphanum}
\alpha
&=&2\int^\infty_{r_0} \frac{bdr}{\sqrt{r^4-b^2r^2+2Mb^2r-b^2Q^2}} -\pi \nonumber\\
&=&2\int^{h_0}_0 \frac{dh}{\sqrt{G(h)}} -\pi, 
\end{eqnarray}
where $h$, $h_0$, and $G(h)$ are defined by $h\equiv M/r$, $h_0\equiv M/r_0$, and 
\begin{eqnarray}
G(h)\equiv -q^2h^4+2h^3-h^2+h_b^2, 
\end{eqnarray}
respectively, and where $q$ and $h_b$ are defined by
$q\equiv Q/M$ and $h_b\equiv M/b$.
By using the relation
\begin{equation}
h_b^2=q^2h_0^4-2h_0^3+h_0^2,
\end{equation}
we can rewrite $G(h)$ as
\begin{eqnarray}
G(h)
&=& (h-h_0)\left[-q^2h^3+(-q^2h_0+2)h^2 \right. \nonumber\\
&& \left. +(-q^2h_0^2+2h_0-1)h+(-q^2h_0^2+2h_0-1)h_0\right] \nonumber\\
&=& -q^2(h-h_0)(h-h_{-1})(h-h_{1})(h-h_{2}),
\end{eqnarray}
where $h_{-1}$, $h_1$, and $h_2$ are three of four real solutions of $G(h)=0$ and 
the relation $h_{-1}<h_0<h_1<h_2$ holds.
The deflection angle can be expressed as
\begin{eqnarray}
\alpha
&=&\frac{2}{q}\int^{h_0}_{h_{-1}} \frac{dh}{\sqrt{-(h-h_0)(h-h_{-1})(h-h_{1})(h-h_{2})}}  \nonumber\\ 
&&-\frac{2}{q}\int^{0}_{h_{-1}} \frac{dh}{\sqrt{-(h-h_0)(h-h_{-1})(h-h_{1})(h-h_{2})}} -\pi  \nonumber\\ 
&=&\frac{4}{q\sqrt{(h_2-h_0)(h_1-h_{-1})}} \left[ K(k)-F\left( \Psi, k \right)  \right] -\pi,  
\end{eqnarray}
where $F\left( \Psi, k \right)$ is the incomplete elliptic integral of the first kind defined by 
\begin{equation}
F(\Psi, k)\equiv \int^{\Psi}_{0}\frac{d\vartheta}{\sqrt{1-k^{2}\sin^{2}\vartheta}},
\end{equation}
$K(k)$ is the complete elliptic integral of the first kind defined by
$K(k)\equiv F(\pi/2,k)$,
$k$ is defined by
\begin{equation}
k \equiv \sqrt{\frac{(h_2-h_1)(h_0-h_{-1})}{(h_2-h_0)(h_1-h_{-1})}},
\end{equation}
and $\Psi$ is defined by
\begin{equation}
\Psi \equiv \arcsin \sqrt{\frac{(h_0-h_2)h_{-1}}{(h_0-h_{-1})h_2}}.
\end{equation}

We define variables $k_\mathrm{p}$ and $h_\mathrm{p}$ by
\begin{equation}
k_\mathrm{p} \equiv \sqrt{1-k^2}
\end{equation}
and
\begin{equation}
h_\mathrm{p} \equiv 1- \frac{h_0}{h_\mathrm{m}} = 1-\frac{r_\mathrm{m}}{r_0}= 1-\frac{3+\sqrt{9-8q^2}}{2}h_0,
\end{equation}
respectively, and where $h_\mathrm{m}$ is defined by
\begin{equation}
h_\mathrm{m} \equiv \frac{M}{r_\mathrm{m}}= \frac{2}{3+\sqrt{9-8q^2}}.
\end{equation}
In the strong deflection limit $r_0 \rightarrow r_\mathrm{m}+0$ or $b\rightarrow b_\mathrm{m}+0$,
we obtain $k \rightarrow 1-0$, $k_\mathrm{p}\rightarrow +0$, $h_\mathrm{p}\rightarrow +0$, and $b_\mathrm{p}\rightarrow +0$. 
First, we expand the deflection angle~$\alpha$ in powers of $k_\mathrm{p}$ around $0$; next we expand it in powers of $h_\mathrm{p}$ around $0$;
then we substitute $h_\mathrm{p}$, which is expanded in powers of $b_\mathrm{p}$ around $0$, into the deflection angle~$\alpha$;
and finally, we obtain the affine perturbation series of the deflection angle~$\alpha$ for $0\leq Q< 3M/(2\sqrt{2})$. 
We can obtain the affine perturbation series of the deflection angle~$\alpha$,
but we do not show its general and explicit formula for $0\leq Q< 3M/(2\sqrt{2})$ on this paper 
because it would have an extremely complicated form and because it is generally not easy to simplify it.

\subsection{Extreme charged black hole with $Q=M$}
\begin{figure}[htbp]
\begin{center}
\includegraphics[width=80mm]{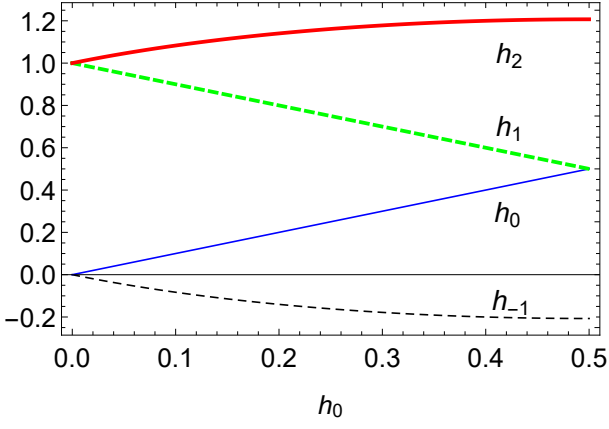}
\end{center}
\caption{$h_2$, $h_1$, $h_0$, and $h_{-1}$
are denoted by wide (red) solid,
wide (green) dashed, 
narrow (blue) solid,  
and narrow (black) dashed curves, respectively,
as a function of $h_0$ in the extreme charged case with $Q=M$.
}
\label{fig:0}
\end{figure}

We concentrate on the extreme charged case with $Q=M$ to show the explicit and very simple form of the affine perturbation series of the deflection angle. 
In this case, we get 
$h_\mathrm{m}=1/2$,
$h_\mathrm{p}=1-2h_0$,
$h_{-1}=(1-\sqrt{1-4 (h_0-1) h_0})/2$, 
$h_1=1-h_0$,
and
$h_2=\left[1+\sqrt{1-4 (h_0-1) h_0}\right]/2$, which are plotted in Fig.~\ref{fig:0}.
By using them and the relation $h_\mathrm{p}=\sqrt{b_\mathrm{p}}$, 
we obtain the affine perturbation series of the deflection angle~$\alpha$ as 
\begin{eqnarray}\label{eq:alpha3}
\alpha&=&-\pi +\frac{2-3 \sqrt{2}}{4}b_\mathrm{p} +\frac{108-149 \sqrt{2}}{256}b_\mathrm{p}^2 \nonumber\\
&&+\frac{2060-2763 \sqrt{2}}{6144}b_\mathrm{p}^3 +O(b_\mathrm{p}^4)\nonumber\\
&&+\frac{2048+768 b_\mathrm{p}+456 b_\mathrm{p}^2+315 b_\mathrm{p}^3 +O(b_\mathrm{p}^4)}{1024 \sqrt{2}} \nonumber\\
&&\times \log \left[ \frac{2^5(3-2\sqrt{2})}{ b_\mathrm{p}} \right] 
\end{eqnarray}
By comparing  Eq.~(\ref{eq:alpha3}) with  Eq.~(\ref{eq:al}), we get 
\begin{eqnarray}
&&\rho_0=-\pi, \; \;
\rho_1=\frac{2-3 \sqrt{2}}{4}, \; \;
\rho_2=\frac{108-149 \sqrt{2}}{256}, \nonumber\\
&&\rho_3=\frac{2060-2763 \sqrt{2}}{6144}, \; \;
\sigma_0=\sqrt{2}, \; \;
\sigma_1=\frac{3\sqrt{2}}{8}, \nonumber\\
&&\sigma_2=\frac{57\sqrt{2}}{256},  \; \; 
\sigma_3=\frac{315\sqrt{2}}{2048}, \;\;
\lambda_0= 2^5(3-2\sqrt{2}).\;\;\;\;
\end{eqnarray}

\subsection{Schwarzschild black hole with $Q=0$}
As a reference, we show the coefficients of the affine perturbation series of the deflection angle in Eq.~(\ref{eq:al}) in the Schwarzschild black hole case with $Q=0$
obtained by Iyer and Petters~\cite{Iyer:2006cn}:
\begin{eqnarray}
&&\rho_0=-\pi, \; \;
\rho_1=\frac{-17+4\sqrt{3}}{18}, \; \;
\rho_2=\frac{-879+236\sqrt{3}}{1296}, \nonumber\\
&&\rho_3=\frac{-321590+90588\sqrt{3}}{629856}, \; \;
\sigma_0=1, \; \;
\sigma_1=\frac{5}{18}, \; \; \nonumber\\ 
&&\sigma_2=\frac{205}{1296}, \;\; 
\sigma_3=\frac{68145}{629856}, \;\;
\lambda_0=216(7-4\sqrt{3}). \;\; \;\;
\end{eqnarray}

\section{Discussion}
If the relation 
\begin{eqnarray}\label{eq:apro}
\lim_{b\rightarrow b_\mathrm{m}+0} \left( \frac{b}{b_\mathrm{m}}-1 \right) = \lim_{b\rightarrow b_\mathrm{m}+0} \left(1- \frac{b_\mathrm{m}}{b} \right)
\end{eqnarray}
is used,
the $0$th order of the affine perturbation series $\alpha_\mathrm{0th}$ in the Schwarzschild spacetime obtained by Iyer and Petters~\cite{Iyer:2006cn} 
corresponds with the deflection angle $\alpha_\mathrm{SDL}$ in the strong deflection limit $r_0 \rightarrow r_\mathrm{m}+0$ or $b\rightarrow b_\mathrm{m}+0$ 
obtained by Darwin~\cite{Darwin_1959} and by Bozza~\cite{Bozza:2002zj}.
From the correspondence, we can express $\bar{a}$ and $\bar{b}$ as
\begin{eqnarray}\label{eq:bara}
\bar{a}= \sigma_0
\end{eqnarray}
and
\begin{eqnarray}\label{eq:barb}
\bar{b}&=&\sigma_0\log \lambda_0 + \rho_0,
\end{eqnarray}
respectively,

In the Reissner-Nordstr\"{o}m spacetime for $0\leq Q< 3M/(2\sqrt{2})$, 
the analytic form of $\bar{a}$ in deflection angle~$\alpha_\mathrm{SDL}$ in the strong deflection limit $b\rightarrow b_\mathrm{m}+0$ was obtained in Refs.~\cite{Bozza:2002zj,Tsukamoto:2016oca,Tsukamoto:2016jzh} as  
\begin{eqnarray}\label{eq:a}
\bar{a} = \frac{r_\mathrm{m}}{\sqrt{3Mr_\mathrm{m}-4Q^2}},
\end{eqnarray}
and $\bar{b}$ was obtained in Refs.~\cite{Tsukamoto:2016oca,Tsukamoto:2016jzh} as
\begin{eqnarray}
\bar{b} 
&=&\bar{a} \log \left[ \frac{8\left(3Mr_\mathrm{m}-4Q^2 \right)^3}{M^2r_\mathrm{m}^2\left(Mr_\mathrm{m}-Q^2 \right)^2} \right. \nonumber\\
&&\left. \times \left( 2\sqrt{Mr_\mathrm{m}-Q^2}-\sqrt{3Mr_\mathrm{m}-4Q^2} \right)^2 \right] -\pi. \nonumber\\
\end{eqnarray}

Therefore, we can obtain the affine perturbation series of the deflection angle in the $0$th order~$\alpha_\mathrm{0th}$~(\ref{eq:0th}) for $0\leq Q< 3M/(2\sqrt{2})$
with 
\begin{eqnarray}
\sigma_0 = \frac{r_\mathrm{m}}{\sqrt{3Mr_\mathrm{m}-4Q^2}},
\end{eqnarray}
\begin{eqnarray}
\lambda_0 
&=& \frac{8\left(3Mr_\mathrm{m}-4Q^2 \right)^3}{M^2r_\mathrm{m}^2\left(Mr_\mathrm{m}-Q^2 \right)^2} \nonumber\\
&&\times \left( 2\sqrt{Mr_\mathrm{m}-Q^2}-\sqrt{3Mr_\mathrm{m}-4Q^2} \right)^2,
\end{eqnarray}
and
\begin{eqnarray}
\rho_0 = -\pi.
\end{eqnarray}

\begin{figure*}[htbp]
\begin{center}
\includegraphics[width=80mm]{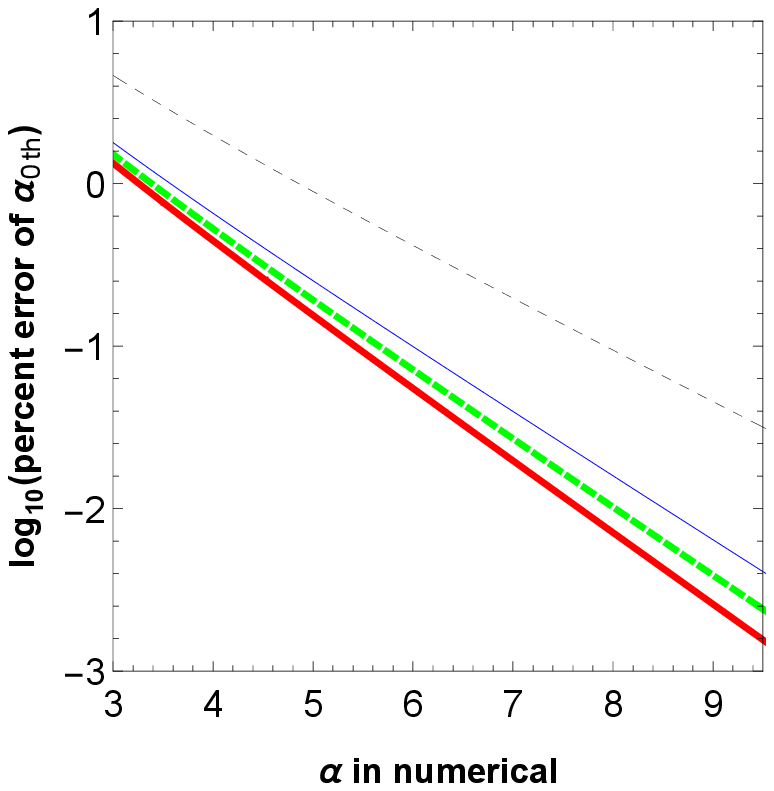}
\includegraphics[width=80mm]{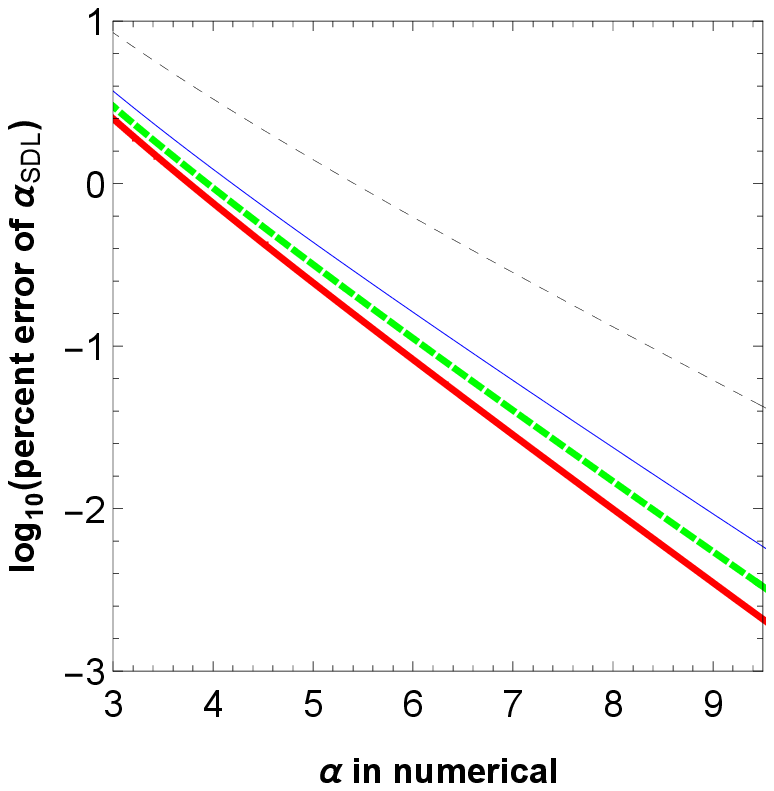}
\end{center}
\caption{The percent errors of the deflection angle in the $0$th order $\alpha_\mathrm{0th}$ (left panel) 
and the deflection angle in the strong deflection limit $\alpha_\mathrm{SDL}$ (right panel) for $Q/M=0$, $0.6$, $0.8$, and $1$ are denoted by
wide (red) solid,
wide (green) dashed, 
narrow (blue) solid,  
and narrow (black) dashed lines, respectively, against $\alpha$ calculated by Eq. (\ref{eq:alphanum}).
}
\label{fig:2}
\end{figure*}

We define the percent errors of $\alpha_\mathrm{0th}$ and $\alpha_\mathrm{SDL}$ as
\begin{eqnarray}
\mathrm{the}\:  \mathrm{percent}\:  \mathrm{error}\:  \mathrm{of}\:  \alpha_\mathrm{0th} \equiv \frac{\alpha-\alpha_\mathrm{0th}}{\alpha} \times 100, 
\end{eqnarray}
and
\begin{eqnarray}
\mathrm{the}\:  \mathrm{percent}\:  \mathrm{error}\:  \mathrm{of}\:  \alpha_\mathrm{SDL} \equiv \frac{\alpha-\alpha_\mathrm{SDL}}{\alpha} \times 100,
\end{eqnarray}
respectively, where $\alpha$ is calculated by using Eq. (\ref{eq:alphanum}),
and the errors are plotted in Fig.~\ref{fig:2}.
We notice that $\alpha_\mathrm{0th}$ has a smaller error than $\alpha_\mathrm{SDL}$. 
The errors of $\alpha_\mathrm{0th}$ and $\alpha_\mathrm{SDL}$ exponentially decrease as $\alpha$ increases.

We also notice that the errors increase as the charge increases and they become very large in the near extreme charged case $Q\sim M$.
This is because the photon sphere exists only for $0\leq Q < 3M/(2\sqrt{2})$ and it becomes a marginally unstable photon sphere with $V_\mathrm{m}=V_\mathrm{m}^{\prime}=V_\mathrm{m}^{\prime \prime}=0$ 
and $V_\mathrm{m}^{\prime \prime \prime}<0$ for $Q=3M/(2\sqrt{2})$ and the deflection angle reflected by the marginally
unstable photon sphere diverges nonlogarithmically~\cite{Tsukamoto:2020iez}.

\begin{figure*}[htbp]
\begin{center}
\includegraphics[width=80mm]{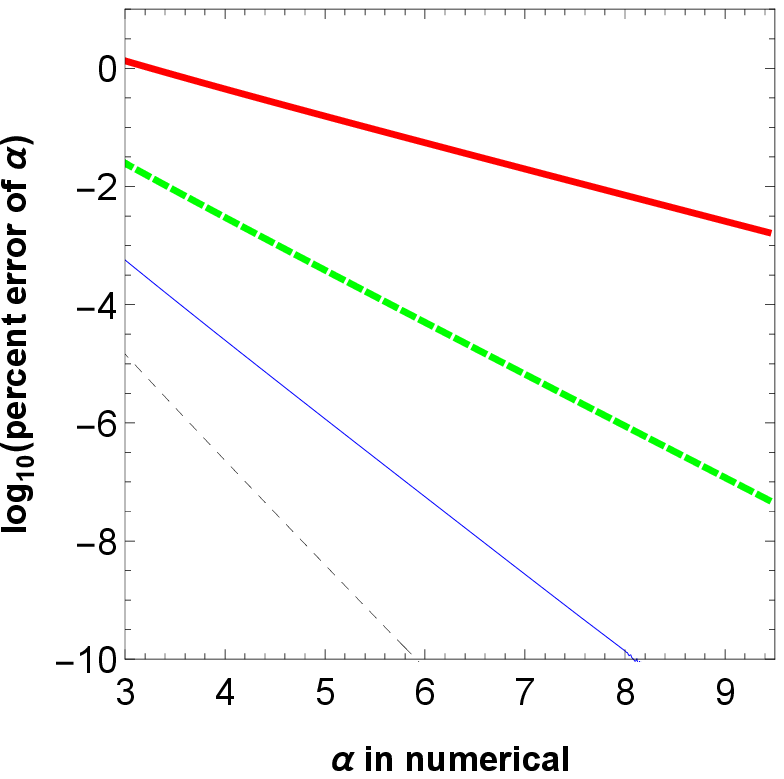}
\includegraphics[width=80mm]{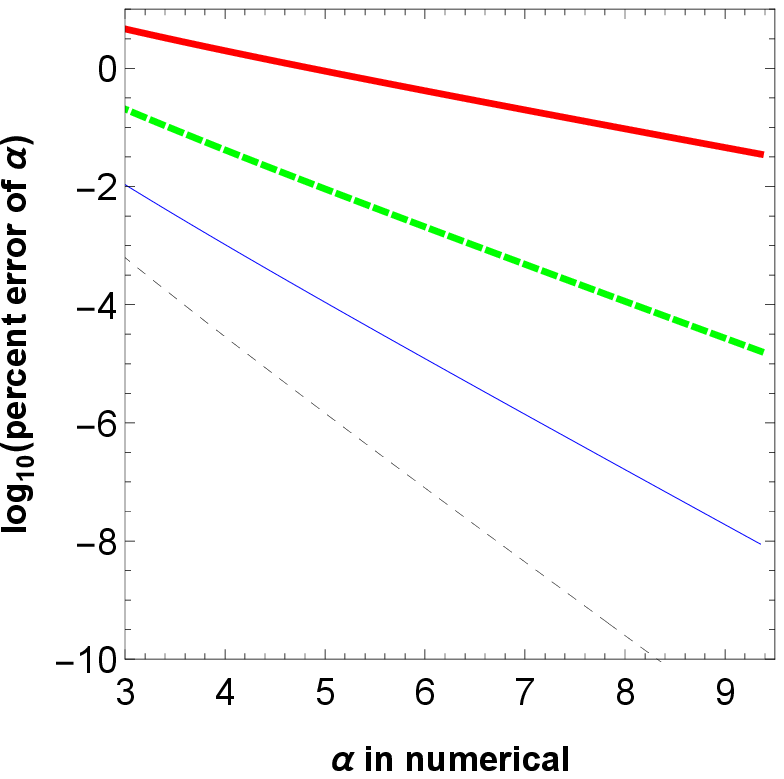}
\end{center}
\caption{The percent errors of $\alpha_\mathrm{0th}$, $\alpha_\mathrm{1st}$, $\alpha_\mathrm{2nd}$, and $\alpha_\mathrm{3rd}$ 
for $Q=0$~(left panel) and $Q=M$~(right panel).
Wide (red) solid,
wide (green) dashed, 
narrow (blue) solid,  
and narrow (black) dashed lines 
denote the percent errors of $\alpha_\mathrm{0th}$, $\alpha_\mathrm{1st}$, $\alpha_\mathrm{2nd}$, and $\alpha_\mathrm{3rd}$, respectively,
against $\alpha$ calculated by using Eq.~(\ref{eq:alphanum}).
}
\label{fig:3}
\end{figure*}

We define the affine perturbation expansion of the deflection angle of the ray in the $1$st order $\alpha_\mathrm{1st}$, the $2$nd order $\alpha_\mathrm{2nd}$, and
the $3$rd order $\alpha_\mathrm{3rd}$ as
\begin{eqnarray}\label{eq:1st}
\alpha_\mathrm{1st}&\equiv&
\left( \sigma_0+\sigma_1 b_\mathrm{p} \right) \log \left( \frac{\lambda_0}{b_\mathrm{p}} \right) +\rho_0+\rho_1 b_\mathrm{p},  \\ \label{eq:2nd}
\alpha_\mathrm{2nd}&\equiv&
\left( \sigma_0+\sigma_1 b_\mathrm{p}+\sigma_2 b_\mathrm{p}^2 \right) \log \left( \frac{\lambda_0}{b_\mathrm{p}} \right) \nonumber\\
&&+\rho_0+\rho_1 b_\mathrm{p}+\rho_2 b_\mathrm{p}^2, 
\end{eqnarray}
and
\begin{eqnarray}\label{eq:3rd}
\alpha_\mathrm{3rd}&\equiv&
\left( \sigma_0+\sigma_1 b_\mathrm{p}+\sigma_2 b_\mathrm{p}^2 +\sigma_3 b_\mathrm{p}^3 \right) \log \left( \frac{\lambda_0}{b_\mathrm{p}} \right) \nonumber\\ 
&&+\rho_0+\rho_1 b_\mathrm{p}+\rho_2 b_\mathrm{p}^2 +\rho_3 b_\mathrm{p}^3,
\end{eqnarray}
respectively.
The percent errors of $\alpha_\mathrm{1st}$, $\alpha_\mathrm{2nd}$, and $\alpha_\mathrm{3rd}$ are defined by 
\begin{eqnarray}
&&\mathrm{the}\:  \mathrm{percent}\:  \mathrm{error}\:  \mathrm{of}\:  \alpha_\mathrm{1st} \equiv \frac{\alpha-\alpha_\mathrm{1st}}{\alpha} \times 100, \\
&&\mathrm{the}\:  \mathrm{percent}\:  \mathrm{error}\:  \mathrm{of}\:  \alpha_\mathrm{2nd} \equiv \frac{\alpha-\alpha_\mathrm{2nd}}{\alpha} \times 100, 
\end{eqnarray}
and
\begin{eqnarray}
\mathrm{the}\:  \mathrm{percent}\:  \mathrm{error}\:  \mathrm{of}\:  \alpha_\mathrm{3rd} \equiv \frac{\alpha-\alpha_\mathrm{3rd}}{\alpha} \times 100, 
\end{eqnarray}
respectively, where $\alpha$ is calculated by Eq. (\ref{eq:alphanum}),
and they are plotted in Fig.~\ref{fig:3}.
\begin{table*}[htbp]
 \caption{Percent errors of $\alpha_\mathrm{SDL}$, $\alpha_\mathrm{0th}$, $\alpha_\mathrm{1th}$, $\alpha_\mathrm{2nd}$, and $\alpha_\mathrm{3rd}$ when $\alpha$ calculated by using Eq. (\ref{eq:alphanum}) is $\pi$, $2\pi$, and $3\pi$
 for the Schwarzschild black hole with $Q=0$. 
 }
\begin{center}
\begin{tabular}{|c||c |c |c |c |c|} \hline
$\alpha$ in numerical &Percent error of $\alpha_\mathrm{SDL}$ &Percent error of $\alpha_\mathrm{0th}$ &Percent error of $\alpha_\mathrm{1th}$ &Percent error of $\alpha_\mathrm{2nd}$ &Percent error of $\alpha_\mathrm{3rd}$\\ \hline
$\pi$ &$2.11$ &$1.14$ &$1.83\times 10^{-2}$ &$3.65\times 10^{-4}$ &$8.09\times 10^{-6}$\\ \hline
$2\pi$ &$6.11\times 10^{-2}$ &$4.11\times 10^{-2}$ &$2.83\times 10^{-5}$ &$2.39\times 10^{-8}$ &$<10^{-10}$\\ \hline
$3\pi$ &$2.26\times 10^{-3}$ &$1.68\times 10^{-3}$ &$5.05\times 10^{-8}$ &$<10^{-10}$ &$<10^{-10}$\\ \hline
\end{tabular}
\end{center}
 \caption{The percent errors in the extreme charged black hole case with $Q=M$.
 }
\begin{center}
\begin{tabular}{|c||c |c |c |c |c|} \hline
$\alpha$ in numerical &Percent error of $\alpha_\mathrm{SDL}$ &Percent error of $\alpha_\mathrm{0th}$ &Percent error of $\alpha_\mathrm{1th}$ &Percent error of $\alpha_\mathrm{2nd}$ &Percent error of $\alpha_\mathrm{3rd}$\\ \hline
$\pi$ &$7.40$  &$4.10$ &$1.62\times 10^{-1}$ &$7.68\times 10^{-3}$ &$4.01\times 10^{-4}$\\ \hline
$2\pi$ &$4.98\times 10^{-1}$ &$3.37\times 10^{-1}$ &$1.37\times 10^{-3}$ &$6.59\times 10^{-6}$ &$3.48\times 10^{-8}$\\ \hline
$3\pi$ &$4.50\times 10^{-2}$ &$3.35\times 10^{-2}$ &$1.47\times 10^{-5}$ &$7.64\times 10^{-9}$ &$<10^{-10}$\\ \hline
\end{tabular}
\end{center}
\end{table*}
Figure~\ref{fig:3} shows that the percent error of the affine perturbation series of the deflection angle in every order decreases exponentially 
as the deflection angle increases and that the percent errors increase if the charge increases. 
The percent errors of $\alpha_\mathrm{SDL}$, $\alpha_\mathrm{0th}$, $\alpha_\mathrm{1st}$, $\alpha_\mathrm{2nd}$, and $\alpha_\mathrm{3rd}$ 
for $\alpha=\pi$, $2\pi$, $3\pi$ are shown in Tables~I and II since they are important in gravitational lensing by the photon sphere of the black hole. 

\begin{figure*}[htbp]
\begin{center}
\includegraphics[width=80mm]{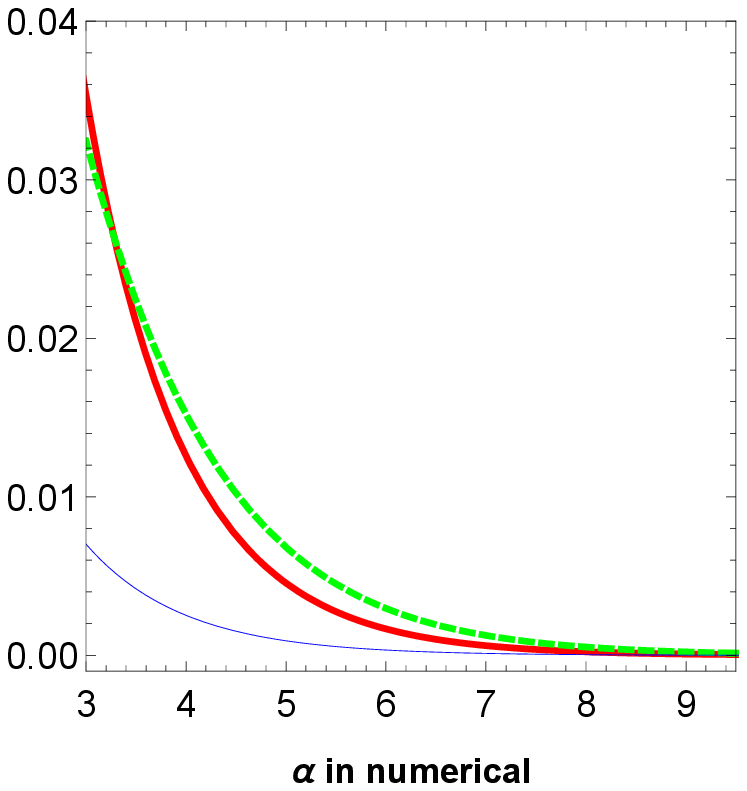}
\includegraphics[width=80mm]{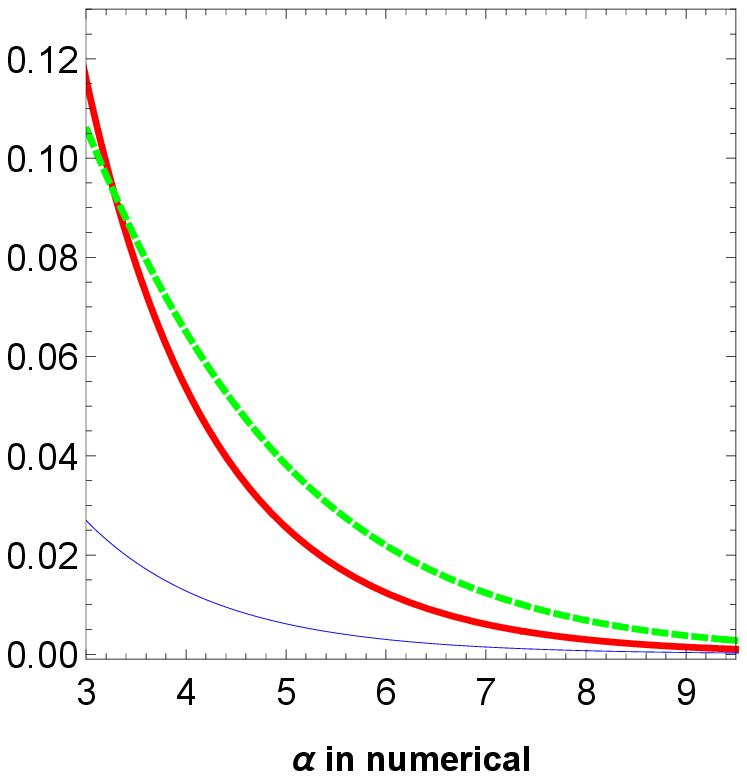}
\end{center}
\caption{The hidden error of $\alpha_\mathrm{SDL}$ given by $\alpha_\mathrm{0th}-\alpha_\mathrm{SDL}=\sigma_0 \log \left[1/(1-b_\mathrm{p})\right]$ and 
the $1$st order terms $-\sigma_1 b_\mathrm{p} \log b_\mathrm{p}$
and $b_\mathrm{p} \left( \rho_1 +\sigma_1 \log \lambda_0 \right)$
are denoted by wide (red) solid,
wide (green) dashed, and
narrow (blue) solid curves, respectively, 
against $\alpha$ calculated by using Eq. (\ref{eq:alphanum})
for $Q=0$~(left panel) and $Q=M$~(right panel).
}
\label{fig:4}
\end{figure*}

By comparing with the errors of $\alpha_\mathrm{0th}$ and $\alpha_\mathrm{SDL}$,
we notice that $\alpha_\mathrm{SDL}$ has a hidden error which is given by a difference
\begin{eqnarray}
\alpha_\mathrm{0th}-\alpha_\mathrm{SDL}=\bar{a}\log\frac{b}{b_\mathrm{m}}=\sigma_0 \log \left(\frac{1}{1-b_\mathrm{p}}\right).
\end{eqnarray}
If the hidden error
is much smaller than the $1$st order terms~$O(b_\mathrm{p}\log b_\mathrm{p})=-\sigma_1 b_\mathrm{p} \log b_\mathrm{p}$ and $O(b_\mathrm{p})=b_\mathrm{p} \left( \rho_1 +\sigma_1 \log \lambda_0 \right)$, 
we do not have to care about the hidden error term.
Note that the higher order terms have been obtained from 
\begin{eqnarray}
\alpha_\mathrm{1st}-\alpha_\mathrm{0th}
&=&\sigma_1 b_\mathrm{p} \log\frac{\lambda_0}{b_\mathrm{p}}+\rho_1 b_\mathrm{p} \nonumber\\
&=&-\sigma_1  b_\mathrm{p} \log b_\mathrm{p} +  b_\mathrm{p} \left( \rho_1 + \sigma_1 \log \lambda_0 \right). \;\; \;\;\;\; \;\;
\end{eqnarray}
The hidden error $\alpha_\mathrm{0th}-\alpha_\mathrm{SDL}=\sigma_0 \log \left[1/(1-b_\mathrm{p})\right]$ and 
the $1$st order terms $-\sigma_1 b_\mathrm{p} \log b_\mathrm{p}$
and $b_\mathrm{p} \left( \rho_1 +\sigma_1 \log \lambda_0 \right)$ are plotted in Fig.~\ref{fig:4}.
We find that the hidden error of $\alpha_\mathrm{SDL}$ is comparable with the $1$st order terms in a retro lensing configuration with $\alpha\sim \pi$.

\section{Conclusion}
The deflection angle in the $0$th order of the affine perturbation series $\alpha_\mathrm{0th}$
is more accurate than the deflection angle $\alpha_\mathrm{SDL}$ in the strong deflection limit
because $\alpha_\mathrm{SDL}$ has a hidden error.
The hidden error given by $\alpha_\mathrm{0th}-\alpha_\mathrm{SDL}$ is comparable with $1$st order terms when the deflection angle is $\pi$.
This implies that we should improve a strong-deflection-limit analysis for a gravitational lensing by using $\alpha_\mathrm{0th}$.
The improvement would not be difficult since we obtain $\alpha_\mathrm{0th}$ from $\alpha_\mathrm{SDL}$ by using relation (\ref{eq:apro}).

%

\end{document}